\documentclass[twocolumn,showpacs,aps,prd]{revtex4}
\usepackage{xspace}
\usepackage{graphicx}
\usepackage{amsmath}
\usepackage{amssymb}
\usepackage{epsfig}

\def\ndata{4985}
\def\ndatb{409}
\def\ndatc{23}
\def\ndatd{1}

\def\twoaa{4.93\pm 0.02}
\def\twoab{3.04\pm 0.10}
\def\twoba{0.21\pm 0.01}
\def\twobb{2.65\pm 0.09}

\def\ba{(2.31\pm 0.04\pm 0.08)\times 10^{-4}}
\def\bb{(1.60\pm 0.20\pm 0.22)\times 10^{-5}}

\def\bka{98\pm 17}
\def\bkb{35\pm 7}

\def\systeffa{0.008}
\def\systbkgda{0.004}
\def\systcomma{0.034}
\def\systalla{0.035}
\def\systeffb{0.12}
\def\systbkgdb{0.04}
\def\systcommb{0.03}
\def\systallb{0.13}

\def\correl{-0.21}

\def\brkkk{(1.9\pm 3.0\pm 0.3)\times 10^{-7}}
\def\brkkklimit{6.3\times 10^{-7}}
\def\brkkkbkgd{20.0\pm 0.5}
\def\kkkeffcc{3.85\pm 0.04}

\def\brkkkz{(1.5\pm 1.8\pm 0.1)\times 10^{-7}}
\def\brkkkzlimit{4.0\times 10^{-7}}
\def\brkkkzbkgd{0.15\pm 0.02}
\def\kkkeffdd{1.37\pm 0.03}

\include{babarsym}

\def\taukstarkz{\ensuremath{\taum\!\rightarrow\Kstarm\Kz \nut}\xspace }

\def\taukstarkzpim{\ensuremath{\taum\!\rightarrow\Kstarz\Kz\pim \nut}\xspace }

\def\taukstarkzpz{\ensuremath{\taum\!\rightarrow\Kstarm\Kz\piz \nut}\xspace }

\def\taupkzkz{\ensuremath{\taum\!\rightarrow\pim\Kz\Kzb \nut}\xspace }

\def\taupkk{\ensuremath{\taum\!\rightarrow\pim\KS\KS \nut}\xspace }
\def\taupkkz{\ensuremath{\taum\!\rightarrow\pim\KS\KS \piz \nut}\xspace }
\def\taupkkg{\ensuremath{\taum\!\rightarrow\pim\KS\KS (\piz) \nut}\xspace }

\def\taukkzkz{\ensuremath{\taum\!\rightarrow\Km\Kz\Kzb \nut}\xspace }

\def\taukkk{\ensuremath{\taum\!\rightarrow\Km\KS\KS \nut}\xspace }
\def\taukkkz{\ensuremath{\taum\!\rightarrow\Km\KS\KS\piz \nut}\xspace }
\def\taukkkg{\ensuremath{\taum\!\rightarrow\Km\KS\KS (\piz) \nut}\xspace }

\def\taupkskl{\ensuremath{\taum\!\rightarrow\pim\KS\KL \nut}\xspace }
\def\taupklkl{\ensuremath{\taum\!\rightarrow\pim\KL\KL \nut}\xspace }

\def\taupkpkm{\ensuremath{\taum\!\rightarrow\pim\Kp\Km \nut}\xspace }
\def\taukkpkm{\ensuremath{\taum\!\rightarrow\Km\Kp\Km \nut}\xspace }

\def\kspp{\ensuremath{\KS\!\rightarrow\pip\pim}\xspace }

\def\taukphi{\ensuremath{\taum\!\rightarrow\Km\phi \nut}\xspace }

\begin{document}
%
\author{J.~P.~Lees}
\author{V.~Poireau}
\author{V.~Tisserand}
\affiliation{Laboratoire d'Annecy-le-Vieux de Physique des Particules (LAPP), Universit\'e de Savoie, CNRS/IN2P3,  F-74941 Annecy-Le-Vieux, France}
\author{J.~Garra~Tico}
\author{E.~Grauges}
\affiliation{Universitat de Barcelona, Facultat de Fisica, Departament ECM, E-08028 Barcelona, Spain }
\author{A.~Palano$^{ab}$ }
\affiliation{INFN Sezione di Bari$^{a}$; Dipartimento di Fisica, Universit\`a di Bari$^{b}$, I-70126 Bari, Italy }
\author{G.~Eigen}
\author{B.~Stugu}
\affiliation{University of Bergen, Institute of Physics, N-5007 Bergen, Norway }
\author{D.~N.~Brown}
\author{L.~T.~Kerth}
\author{Yu.~G.~Kolomensky}
\author{G.~Lynch}
\affiliation{Lawrence Berkeley National Laboratory and University of California, Berkeley, California 94720, USA }
% ...
\collaboration{The \babar\ Collaboration}
\noaffiliation

\title{The branching fraction of
{\boldmath $\taum \rightarrow \pim \KS \KS \, (\piz) \nut$} decays }

\begin{flushleft}
BaBar-PUB-12/021 \\
SLAC-PUB-15206 \\
\end{flushleft}


\vspace{1pc}
\begin{abstract}
\begin{center}
\large \bf Abstract
\end{center}
We present a study of \taupkkg and \taukkkg
decays using a dataset of 430 million $\tau$ lepton pairs, 
corresponding to an integrated luminosity of $468\,\invfb$, collected with the 
\babar\ detector at the \pep2\ asymmetric energy \epem storage rings.  
We measure branching fractions of $\ba$ and $\bb$ for the \taupkk
and \taupkkz decays, respectively.
We find no evidence for \taukkk and \taukkkz decays and place 
upper limits on the branching fractions of $\brkkklimit$ and $\brkkkzlimit$ 
at the 90\% confidence level.
\vspace{1pc}
\end{abstract}


\maketitle


The \mtau lepton can be used as a high-precision probe of 
the Standard Model (SM) and models of new physics.
A recent \babar\ paper, for example,  presented a search for \CP 
violation by measuring the decay-rate asymmetry of 
$\taum \rightarrow \pim \KS \nut$ decays \cite{babar:cpv}.
One of the backgrounds in that analysis is \taupkk, 
which has a large uncertainty in the branching fraction \cite{pdg}.
The uncertainty in the background from \taupkk decays was not a limitation 
of the decay-rate asymmetry measurement, but an improved measurement
of the branching fraction and an understanding of the decay dynamics will 
be required for a future measurement at a high-luminosity $B$-factory.

This paper presents measurements of the branching fractions of
\taupkkg decays and  the first search for \taukkkg decays.
In this work we use the \kspp decay mode.
Here and throughout the paper, charge conjugation is implied.

Previously, ALEPH and CLEO measured the \taupkk branching fraction to be
$(2.6 \pm 1.0 \pm 0.5) \times 10^{-4}$ \cite{aleph}
and  $(2.3 \pm 0.5 \pm 0.3) \times 10^{-4}$ \cite{cleo}, respectively.
ALEPH set an upper limit on the \taupkkz branching fraction of 
$2\times 10^{-4}$  at the 95\% confidence level \cite{aleph}.

The present analysis uses data recorded by the \babar\ detector at the \pep2\ 
asymmetric-energy \epem\ collider, operated at center-of-mass (CM) 
energies of 10.58\gev and 10.54\gev at the SLAC National Accelerator Laboratory.
The \babar\ detector is described in detail in Ref.~\cite{detector}.  
In particular, charged particle momenta are measured with a five-layer
double-sided silicon vertex tracker and a 40-layer drift chamber, both within 
a 1.5 T superconducting solenoidal magnet. 
Charged kaons and pions are separated by ionization ($dE/dx$) 
measurements in the silicon vertex detector and the drift chamber
in combination with an internally reflecting Cherenkov detector.
An electromagnetic calorimeter made of thallium-doped cesium iodide crystals 
provides energy measurements for electrons and photons, 
and an instrumented flux return detector identifies muons. 
Based on an integrated luminosity of 468\invfb, 
the data sample contains approximately 430 million \mtau-pair events.

Simulated event samples are used to estimate the selection efficiency 
and purity of the data sample.  
The production of \mtau pairs is simulated with the KK2F Monte Carlo (MC)
event generator \cite{kk}.  
Subsequent decays of the $\tau$ lepton, 
continuum \qqbar events (where $q=u,d,s,c$), 
and final-state radiative effects are modeled with Tauola \cite{tauola} 
and EvtGen \cite{evtgen},
JETSET \cite{jetset}, and PHOTOS \cite{photos}, respectively.  
Passage of the particles through the detector is simulated by Geant4 
\cite{geant}.

The \taupkk decay is simulated with Tauola using \taukstarkz. 
The \taupkkz decay is simulated with EvtGen using \taukstarkzpz 
and \taukstarkzpim.
As we later show, the \taukstarkz and \taukstarkzpz have a $\Kstar(892)$ 
meson that is observed in the $\pim\KS$ channel, and the \taukstarkzpim 
has a $\Kstar(892)$ meson that is observed in $\piz\KS$ channel.

The $\tau$ pair is produced back-to-back in the \epem CM frame.
As a result, the decay products of the two $\tau$ leptons can be separated
from each other by dividing the event into two hemispheres -- referred 
later as the ``signal'' hemisphere and the ``tag'' hemisphere -- using 
the plane perpendicular to the event thrust axis \cite{thrust}.  
The event thrust axis is calculated using 
all charged particles and all photon candidates in the entire event.

\begin{figure}[hb] 
\begin{center}
\mbox{\epsfig{file=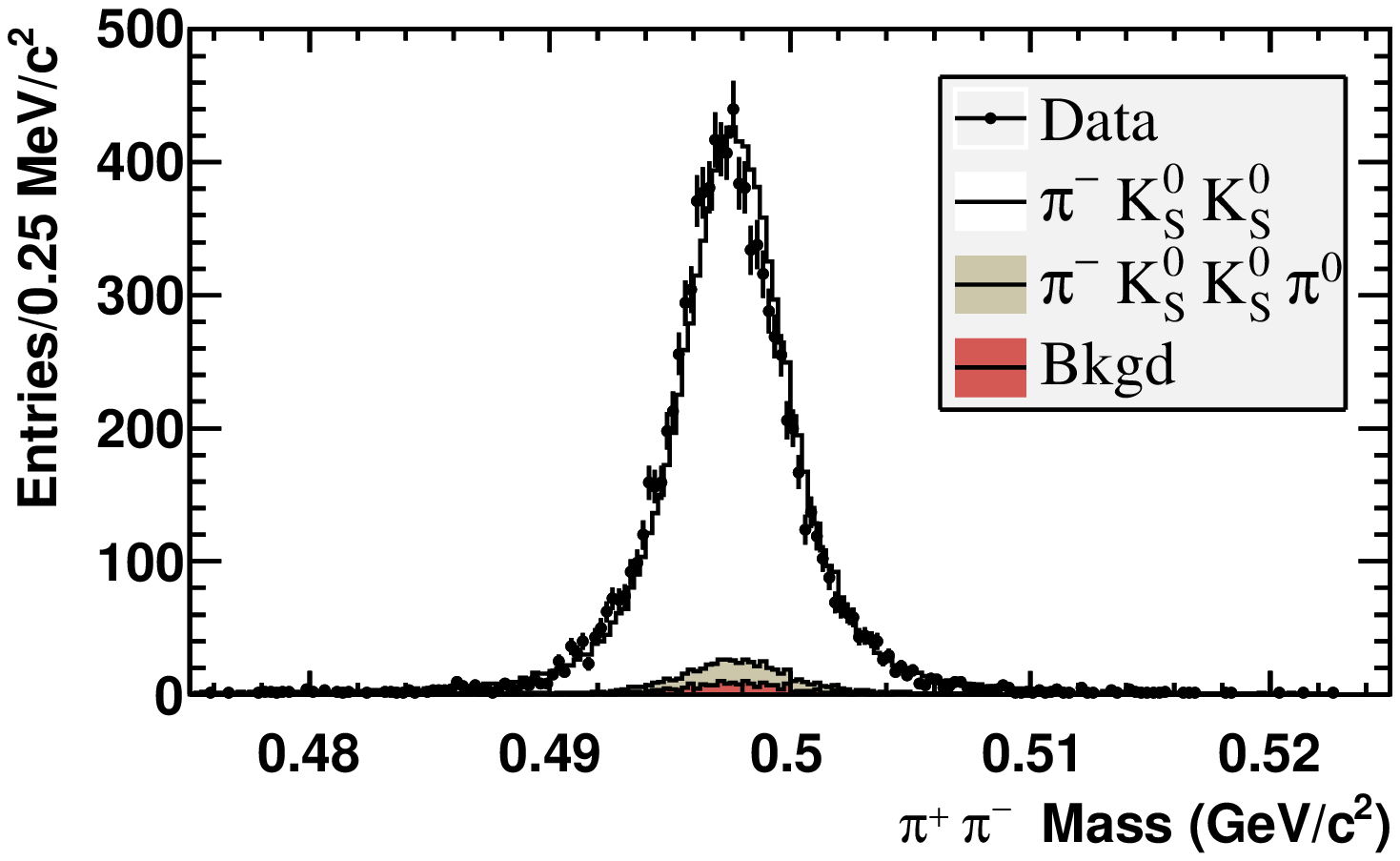,height=5.5cm}} 
\mbox{\epsfig{file=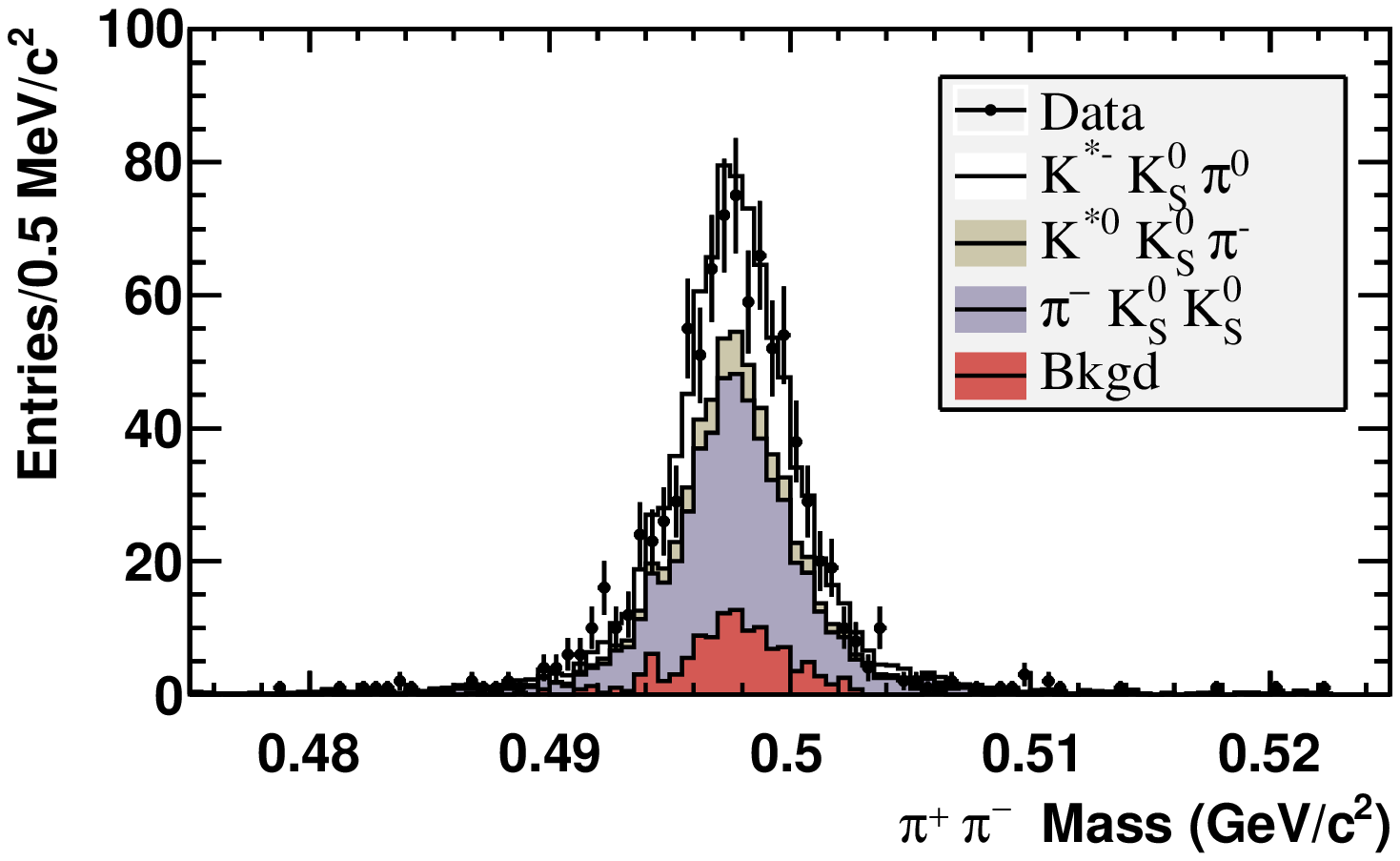,height=5.5cm}} 
\end{center}
\caption{\label{fig:ksselection}
The invariant mass of the two $\KS \rightarrow \pip \pim$ candidates 
in the \taupkk (top) and \taupkkz (bottom) samples
after all selection criteria have been applied.
The points are data and the histograms are the prediction
of the Monte Carlo simulation.
For both plots, the white histogram represents \taukstarkz decays,
the blue and beige histogram shows the \taukstarkzpz and \taukstarkzpim 
(\taupkkz) decays, respectively.
The red histogram is the \qqbar background.
}
\end{figure}

We select events with one prompt track and two \KS\to\pip\pim candidates 
reconstructed in the signal hemisphere, 
and exactly one oppositely charged prompt track in the tag hemisphere.  
A prompt track is defined to be a track with 
its point of closest approach to the beam spot being less than 1.5\cm  
in the plane transverse to the \en beam axis and 
less than 2.5\cm in the direction of the \en beam axis.  
Tracks consistent with coming from a \KS or $\Lambda$ decay, 
or from a \g conversion are not considered to be prompt tracks.
The components of momentum transverse to the \en beam axis for 
each of these two prompt tracks must be greater 
than 0.1\gevc in the laboratory frame.  
A \KS candidate is defined as a pair of oppositely charged pion candidates 
with an invariant mass between 0.475 and 0.525\gevcc 
(see Fig.~\ref{fig:ksselection}); 
furthermore, the distance between the beam spot and the \pip\pim vertex 
must be at least three times its uncertainty 
(the di-pion pair will be referred to as the ``\KS candidate daughters'').  

The charged hadron must be identified as a charged pion or a charged kaon.
The efficiency for selecting charged pions and kaons is approximately 
95\% and 90\%, respectively.
The probability of mis-identifying a charged pion (kaon) as a charged 
kaon (pion) is estimated to be 1\% (5\%).

The charged pion and kaon samples are divided into samples
with zero and one \piz mesons.
Events with two or more \piz mesons are rejected.
The \piz candidate is reconstructed from two clusters 
of energy deposits in the electromagnetic calorimeter that have no 
associated tracks.  
The energy of each cluster is required to be greater than 30\mev in 
the laboratory frame, and the invariant mass of the two clusters must be 
between 0.115\gevcc and 0.150\gevcc.  
The clusters in the electromagnetic calorimeter that
are not associated with a \piz candidate
are ignored in the analysis.

To reduce backgrounds from non-\mtau-pair events, we require that the 
momentum of the charged particle in the tag hemisphere is less than 4\gevc 
in the CM frame and be identified as either an electron or a muon.
For momenta above 1\gevc in the laboratory frame, 
electrons and muons are identified with efficiencies of approximately 
92\% and 70\%, respectively \cite{tautolll}. 
We also require the magnitude of the event thrust to be between 0.90 and 0.995.

\begin{figure*}[hbt] 
\begin{center}
\mbox{\epsfig{file=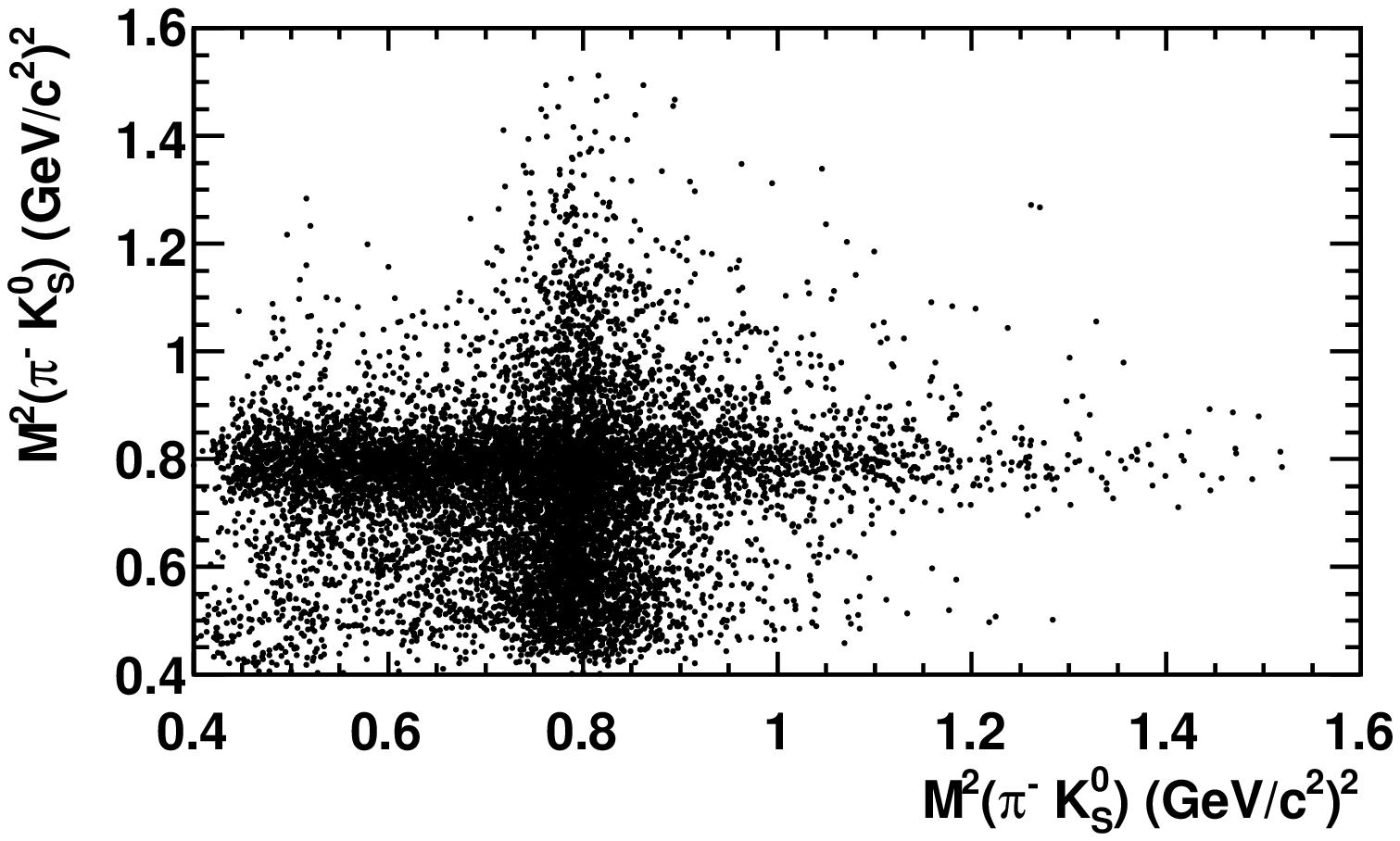,height=5.5cm}} 
\mbox{\epsfig{file=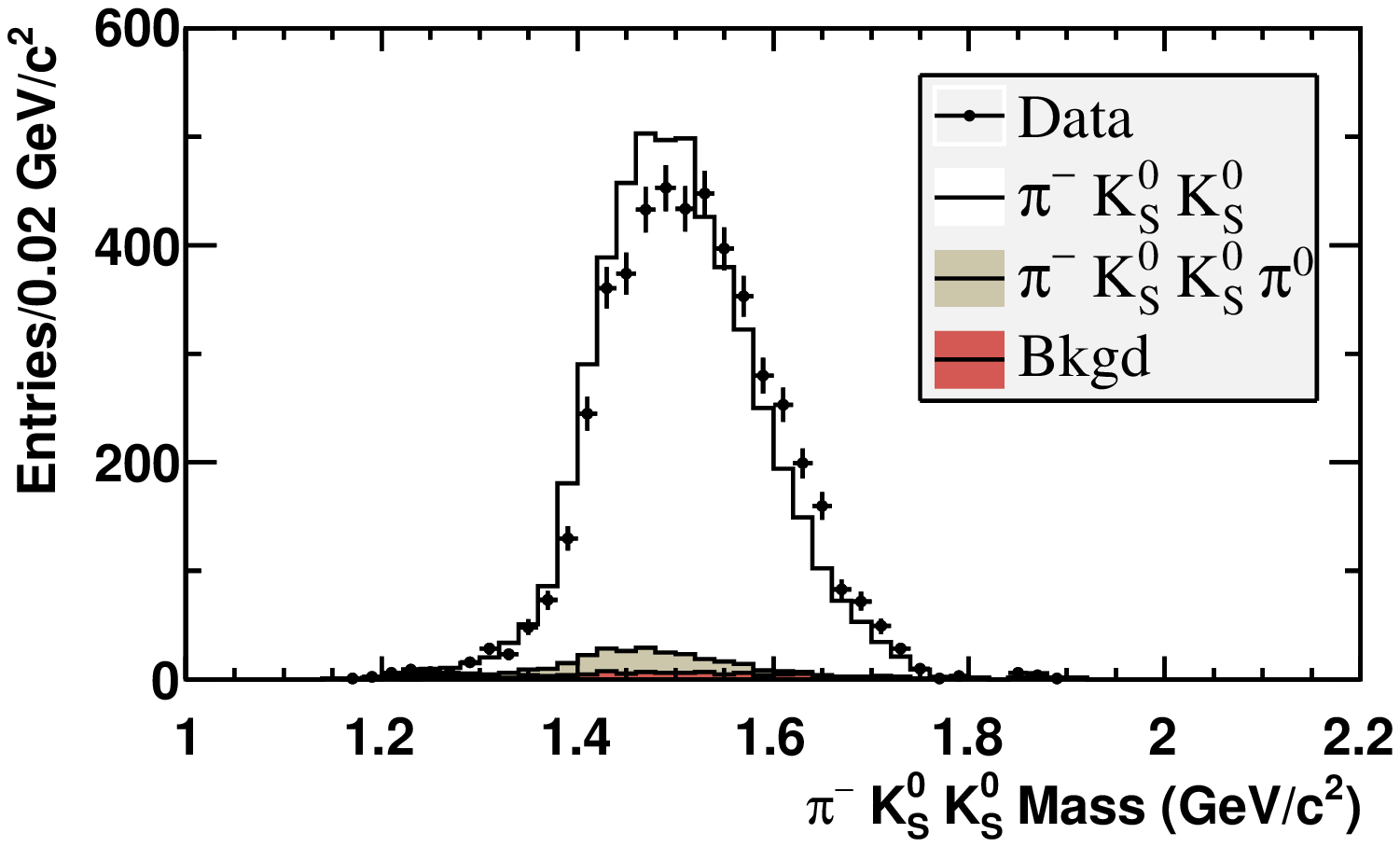,height=5.5cm}} 
\mbox{\epsfig{file=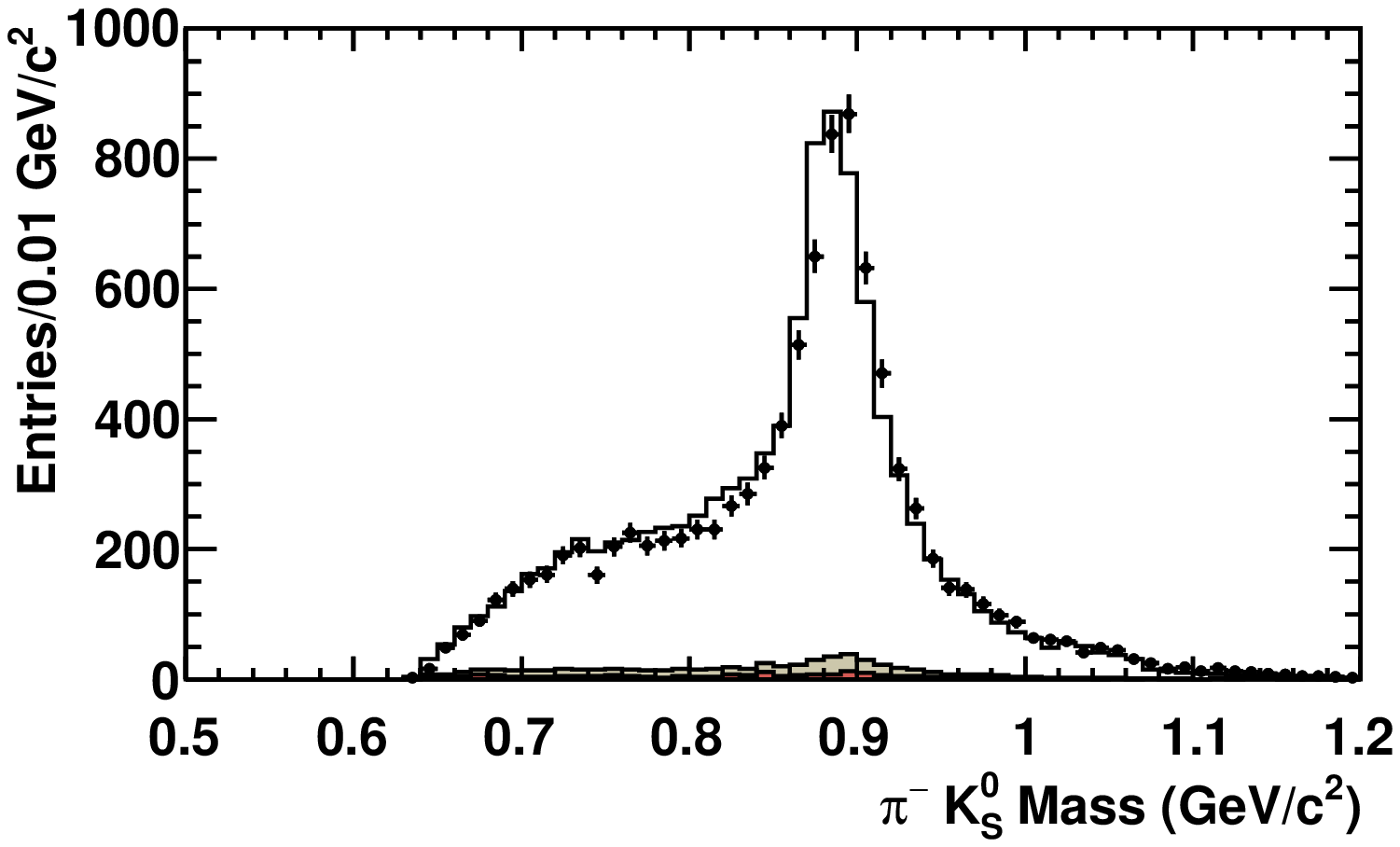,height=5.5cm}} 
\mbox{\epsfig{file=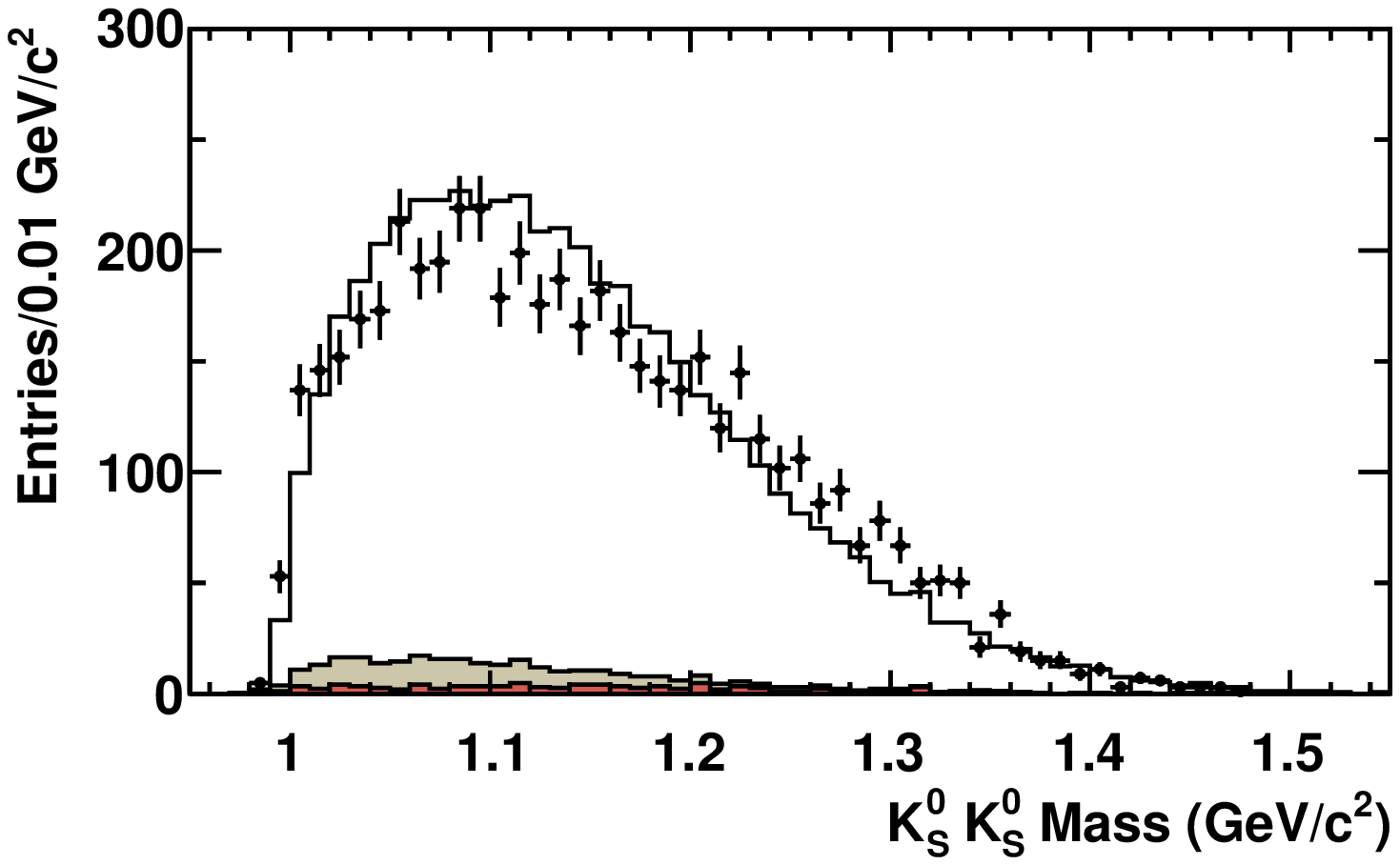,height=5.5cm}} 
\end{center}
\caption{\label{fig:pkk}
The Dalitz plot of the $(\pim\KS)$ system, and the $(\pim\KS\KS)$, 
$(\pim\KS)$ and $(\KS\KS)$ invariant mass distributions 
for events that pass the \taupkk selection criteria.
There are two entries per event in the Dalitz plot.
The points are data and the histograms are the prediction
of the Monte Carlo simulation.
The signal decays are represented by the white histogram
(\taukstarkz).
The beige histogram shows the \taukstarkzpz and \taukstarkzpim 
(\taupkkz) decays,
The red histogram is the \qqbar background.
The mass plots use \taupkk events that have been weighted based on the 
Dalitz plot distributions in the top left plot.
}
\end{figure*}

\begin{figure*}[hbt] 
\begin{center}
\mbox{\epsfig{file=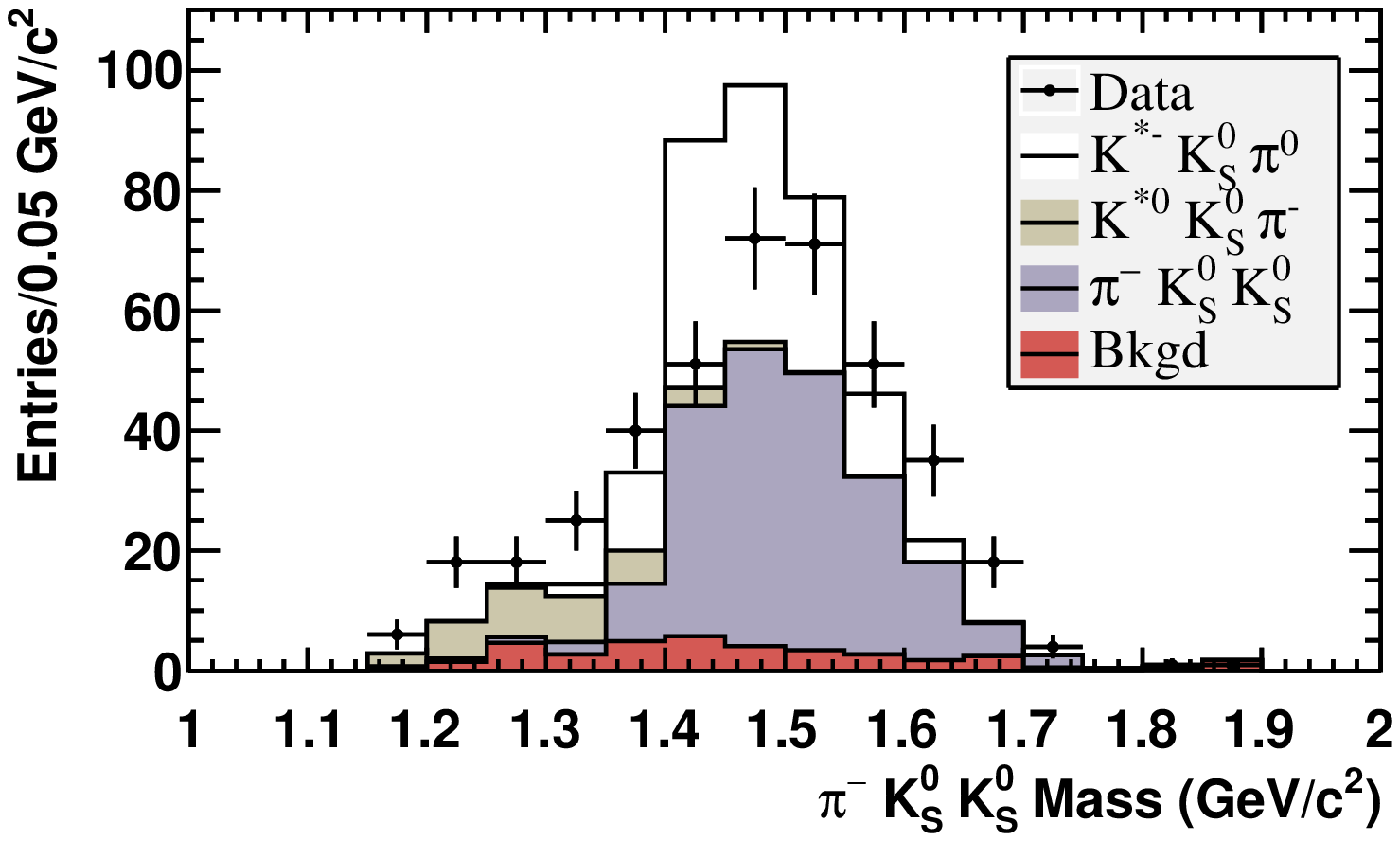,height=5.5cm}} 
\mbox{\epsfig{file=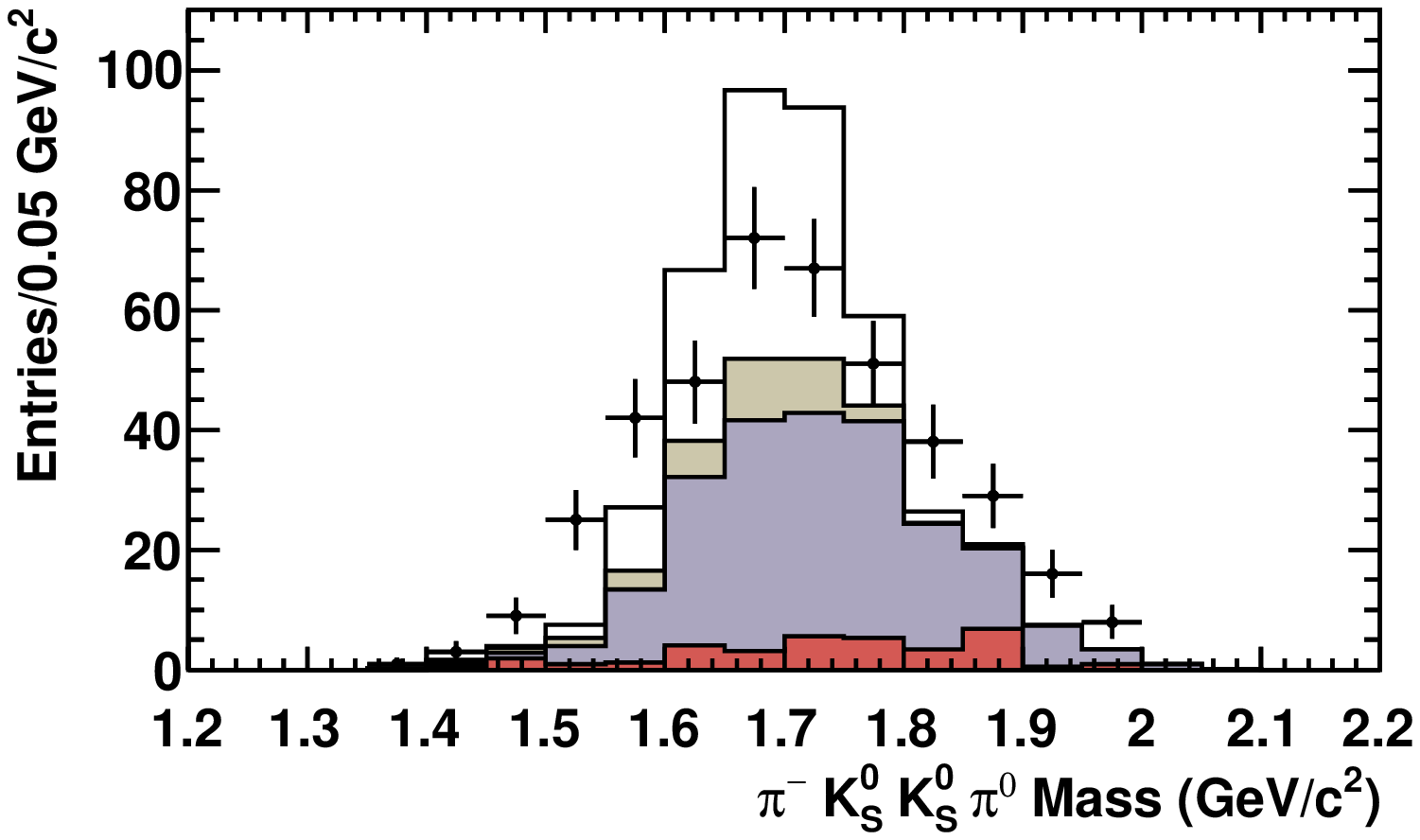,height=5.5cm}} 
\mbox{\epsfig{file=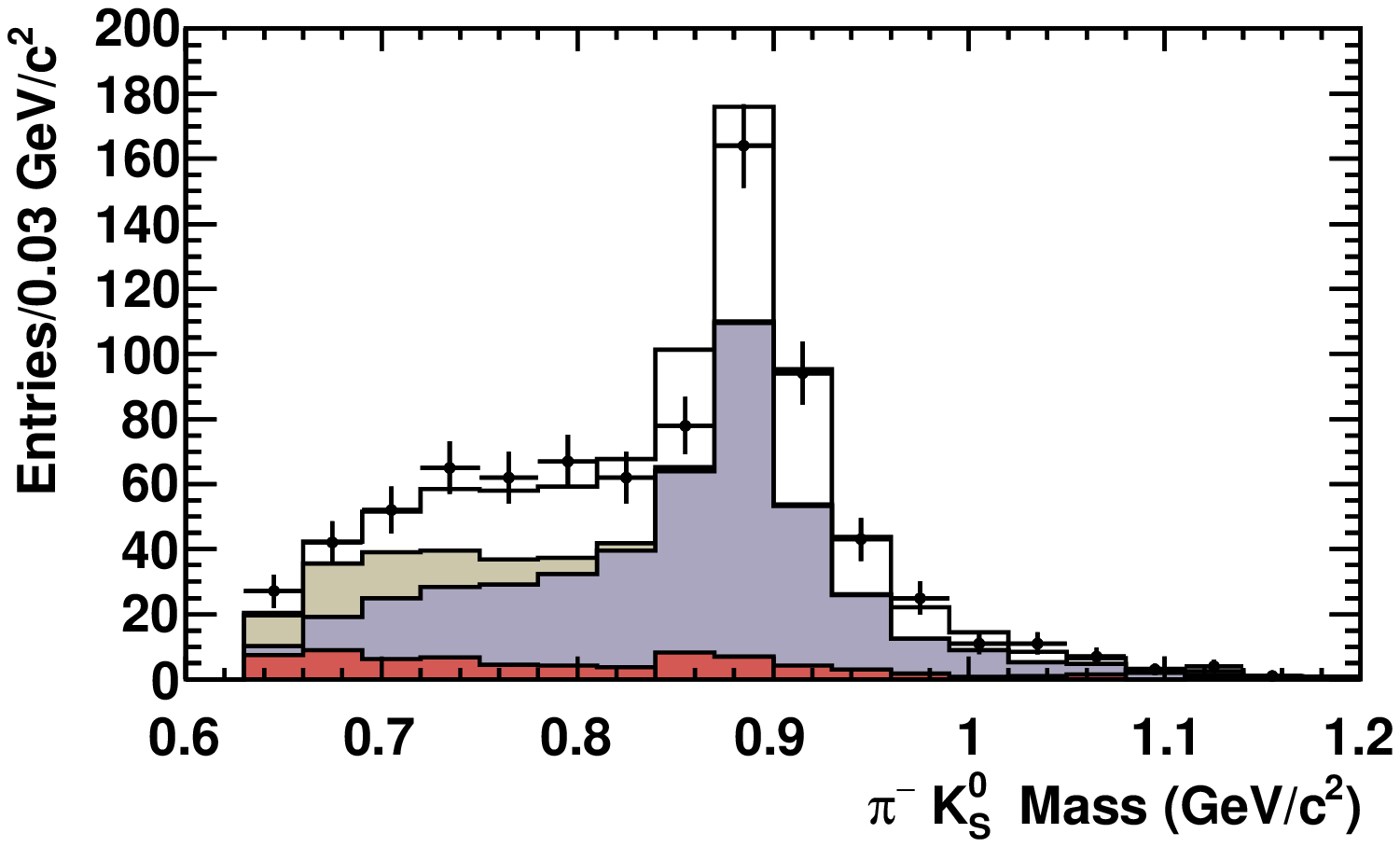,height=5.5cm}} 
\mbox{\epsfig{file=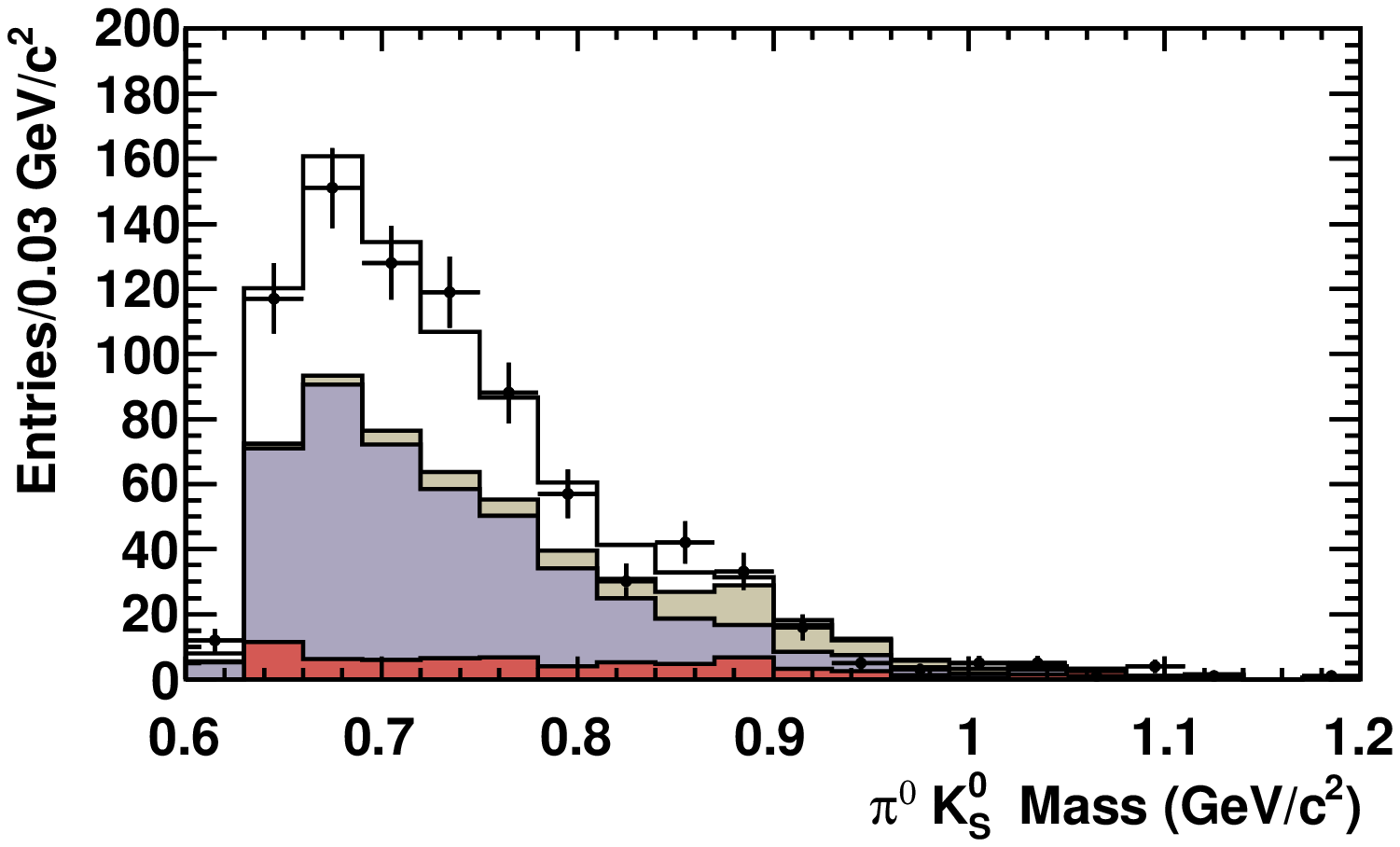,height=5.5cm}} 
\end{center}
\caption{\label{fig:pkkz}
The $(\pim\KS\KS)$, $(\pim\KS\KS\piz)$, $(\pim\KS)$, and $(\piz\KS)$ invariant 
mass distributions that pass the \taupkkz selection criteria.
The points are data and the histograms are the predictions
of the Monte Carlo simulation.
The two signal channels are shown in the white (\taukstarkzpz) 
and beige (\taukstarkzpim) histograms.
The dark blue histogram is \taupkk (\taukstarkz) decays.
The red histogram is the \qqbar background.
The mass plots use \taupkk events that have been weighted based on the
Dalitz plot distributions shown in the Fig.~2.
}
\end{figure*}

The invariant mass of the charged hadron and the two \KS 
mesons is required to be less than 1.8 \gevcc.
For \taupkkz decays, we do not include the \piz in the mass calculation.
The $\pim\KS\KS$ invariant mass is shown in Figs.~\ref{fig:pkk} and 
\ref{fig:pkkz}.   
The  $\pim\KS\KS\piz$ invariant mass is also shown in Fig.~\ref{fig:pkkz}.
We also require that the pseudomass to be less than 1.9 and 2.1 \gevcc
for the \taupkk and \taupkkz samples, respectively (the \piz meson
is included in the pseudomass calculation).
The pseudomass is defined to be 
$M_{\mathrm{pseudo}} = \sqrt{ M_h^2 + 2(\sqrt{s} - E_h)(E_h - P_h)}$
where $E_h$ and $P_h$ are the energy and magnitude of the momentum
of the hadronic final state in the laboratory frame \cite{stahl}.

The invariant mass distribution predicted by the MC for the 
hadronic final state particles and for their combinations 
do not perfectly describe the data.
In particular, the peak of the $(\pim\KS\KS)$ invariant mass
distribution in the MC is found to peak approximately 5\% lower than
the peak observed in the data.
To improve the modeling of the data we have weighted the \taupkk in Tauola
using the Dalitz plot distribution for the $\pim\KS$ invariant mass
(shown for the data sample in Fig.~\ref{fig:pkk}).
The weighting function is from a two-dimensional ($9 \times 9$) matrix 
using $M^2(\pim\KS)$ with both $\pim\KS$ combinations (the matrix
is constructed to be symmetric).
The weighted events are used in all the mass plots and we observe
an improvement in the modeling of the data.

The branching fractions of the two charged pion modes are determined
simultaneously to take into account the cross feed of each decay mode
into the other sample.
The branching fraction is 

\begin{eqnarray*} 
B_j & = \sum_{i} \epsilon_{ji}^{-1} 
(N_i^{\mathrm{data}} - N_i^{\mathrm{bkgd}})  /  (2 N_{\tau\tau})
\end{eqnarray*}


\noindent
where $j$ represents the  \taupkk and \taupkkz decay modes;
$i$ represents the ($\pim\KS\KS$) or ($\pim\KS\KS\piz$) reconstruction modes;
$N_i^{\mathrm{data}}$ and $N_i^{\mathrm{bkgd}}$ are the number of data 
and background events
in the $i$-th data sample;
$\epsilon^{-1}$ is the inverse of the selection efficiency matrix
($\epsilon_{ij}$ is the probability to select an event of
type $j$ with the selection criteria $i$); 
and $N_{\tau\tau}$ is the number of \mtau-pair candidates.

The columns in Table~\ref{table:pion:results} give the number of data 
and background events for each reconstruction mode.
Table~\ref{table:pion:results} also gives the selection efficiency matrix, 
where the horizontal row gives the efficiency for selecting the true decay
for each reconstructed mode.
For example, the efficiency for selecting a true \taupkk decay is 
($\twoaa$)\% and ($\twoba$)\% with the ($\pim\KS\KS$) and 
($\pim\KS\KS\piz$) selection criteria, respectively.

\renewcommand\arraystretch{1.35}
\begin{table*}[htb]
\caption{\label{table:pion:results} Results for the charged pion decays. 
The background events are primarily \qqbar events.}
\begin{center}
\begin{tabular}{lcc}  \hline
Decay mode           & \hspace{1.25cm} \taupkk      \hspace{1.25cm} 
	             & \hspace{1.25cm} \taupkkz  \hspace{1.25cm} \\  \hline 
Branching fraction   & $\ba$        & $\bb$        \\ \hline
Events & & \\
Data                 & \ndata       & \ndatb       \\
Estimated background & $\bka$       & $\bkb$       \\ \hline
Selection efficiency & & \\
\taupkk              & ($\twoaa$)\% & ($\twoba$)\% \\
\taupkkz             & ($\twoab$)\% & ($\twobb$)\% \\
\hline
Fractional systematic errors & & \\
Selection efficiency & \systeffa  & \systeffb \\
Background           & \systbkgda & \systbkgdb \\
Common systematics   & \systcomma & \systcommb \\
Total                & \systalla  & \systallb \\
\hline
\end{tabular}
\end{center}
\end{table*}

We measure the \taupkk and \taupkkz branching fractions 
to be
\begin{eqnarray*}
 B(\taupkk)  =  & \ba \\
 B(\taupkkz) =  & \bb 
\end{eqnarray*}
where the first error is statistical and the second is systematic.
The statistical correlation parameter for the two measurements is found 
to be $\correl$.
The results have been corrected for the \kspp branching fraction \cite{pdg}.

The systematic uncertainties (see Table~\ref{table:pion:results})
are divided into the selection efficiency, background, and common
systematic components.
The uncertainties on the elements of the efficiency matrix only include
the errors specific to that decay and selection criteria.
Uncertainties that are common to all matrix elements are included
in the common systematic errors.

The efficiency for selecting \taupkk events is found to be 
($\twoaa$)\% and ($\twoba$)\% for the samples with zero and one \piz 
candidate, respectively.
The uncertainty on the first efficiency is from the MC statistical
error.
The uncertainty on the second efficiency also includes the MC
statistical error and an error that takes into account the
uncertainty for finding a fake \piz meson in \taupkk decays.
The uncertainty for finding a fake \piz is estimated to be 6\%
and is determined by comparing the number of \taupkk decays that 
have two neutral clusters in the data and MC samples where
the invariant mass of the two neutral clusters must not be near
the \piz mass.

The efficiency for selecting \taupkkz events is found to be 
($\twoab$)\% and ($\twobb$)\% for the samples with zero and one \piz 
candidate, respectively.
The uncertainties include the MC statistical error
and an uncertainty for the \piz identification.
The uncertainty for identifying a \piz meson is estimated to be 3\% 
based on studies involving a variety of data and MC control samples.
We observe that the efficiency for selecting \taupkkz decays with and 
without a \piz is approximately equal, and hence we assign
a 3\% uncertainty on the efficiency for selecting \taupkkz
decays without reconstructing the \piz meson.

The background in the charged pion modes is predicted by the MC
simulation to be entirely from $\epem \rightarrow \qqbar$ events.
The background in the charged kaon modes is cross-feed from the charged 
pion modes where a charged pion is mis-identified as the charged kaon.
The background in the charged pion sample is confirmed with data and 
MC simulation control samples.
The control samples are created using the nominal selection criteria 
except that the invariant mass and pseudomass requirements are reversed to 
eliminate the \mtau-pair events and enhance \qqbar events.
The ratio of selected events in the data to MC control samples is 
found to be consistent with unity within 15\% for both \taupkk 
and \taupkkz samples.
The 15\% value is added to the MC statistical uncertainty
of the number of background events.

A number of systematic uncertainties are common to both the
\taupkk and \taupkkz branching fractions measurements.
They can be categorized into two components: 
tracking and particle identification reconstruction uncertainties,
and topological selection uncertainties.

The tracking and particle identification reconstruction uncertainties
include the uncertainty on the track reconstruction efficiency (0.5\%).
They also include the uncertainties on the efficiencies of the particle 
identification algorithms: lepton identification (combined electron 
and muon) (1.6\%), charged pion particle identification (0.5\%),
and \KS identification (1.8\% for two \KS).
The particle identification algorithms used in this work are
based on standard \babar\ routines and the uncertainties are
determined using control data and MC samples \cite{detector, btohh}.
The uncertainty on the efficiency for selecting \piz mesons is included 
in the elements of the selection efficiency matrix.

The topological selection uncertainties include a 2\% uncertainty 
associated with the topological selection criteria that impose requirements 
that the prompt tracks be associated the primary vertex.   
Also included is the uncertainty in the product of the luminosity 
multiplied by the $\epem \rightarrow \tautau$ cross section (1\%).
The weighting of the invariant mass distribution for
the \taupkk MC decays does not change the number of events and, hence,
does not alter the measured branching fractions.

In Fig.~\ref{fig:pkk} we plot the $(\pim\KS\KS)$, $(\pim\KS)$,
and $(\KS\KS)$ invariant mass distributions.
The contribution of the $\Kstar(892)$ resonance 
($\Kstar \rightarrow \pim \KS$) is observed in the $(\pim\KS)$ invariant 
mass plot and the Dalitz plot in Fig.~\ref{fig:pkk}.

The \taupkk branching fraction is in good agreement with the previous
measurements of $(2.6 \pm 1.0 \pm 0.5) \times 10^{-4}$ \cite{aleph}
and  $(2.3 \pm 0.5 \pm 0.3) \times 10^{-4}$ \cite{cleo}.
The theoretical prediction for the \taupkk  branching fraction is
$4.8 \times 10^{-4}$ \cite{fink}.
Decays involving a pion and two kaon mesons can have contributions
from both axial and vector currents at the same time, and
the vector contribution for \taupkk is estimated to be $1.4 \times 10^{-4}$.

Assuming isospin symmetry \cite{rouge} and using other measurements, we
can estimate the \taupkskl branching fraction.
The \taupkzkz and \taupkpkm  branching fractions 
are equal if isospin is an exact symmetry
(the \taupkk and \taupklkl branching fractions are also equal).
Hence $B(\taupkskl) =  B(\taupkpkm) - 2 B(\taupkk)$ and
we obtain 
\begin{eqnarray*}
B(\taupkskl)  =  & (9.8 \pm  0.5)\times 10^{-4} 
\end{eqnarray*}
where $B(\taupkpkm) = (14.4 \pm 0.4) \times 10^{-4}$ \cite{pdg} based on 
measurements from \babar\ \cite{babar:pkpkm} and Belle \cite{belle:pkpkm}.
The prediction is in good agreement with the branching fraction
measured by ALEPH of $(10.1 \pm 2.3 \pm 1.3) \times 10^{-4}$  \cite{aleph}

The MC simulation uses \taukstarkzpz and \taukstarkzpim decays.
A model based on a phase space distribution of the final state particles
does not describe the $(\pim\KS)$ invariant mass distribution.
The relative contribution of \taukstarkzpz to \taukstarkzpim decays
is determined to be $(0.17 \pm 0.03)$ by simultaneously fitting the 
$(\pim\KS)$ and $(\piz\KS)$ invariant mass distributions 
(see  Fig.~\ref{fig:pkkz}).
The predicted Monte Carlo distributions are fit to the data spectra
after the subtraction of the \taupkk and background events.
The normalizations of the two modes are varied 
with the constraint that the values be positive numbers.
If we do not include the  \taukstarkzpim decay, then we observe a 
disagreement between the data and MC samples in the lower-mass and 
higher-mass regions of the $M(\pim\KS)$ and $M(\piz\KS)$ distributions,
respectively, in Fig.~\ref{fig:pkkz}.

\renewcommand\arraystretch{1.35}
\begin{table*}[htb]
\caption{\label{table:kaon:results} Results for the charged kaon decays.
The background events are primarily \qqbar events.}
\begin{center}
\begin{tabular}{lcc}  \hline
Decay mode            & \hspace{1cm} \taukkk  \hspace{1cm} 
	              & \hspace{1cm} \taukkkz  \hspace{1cm}   \\  \hline
Branching fraction    & $\brkkk$         & $\brkkkz$          \\  
Limit (90\% C.L.)     & $\brkkklimit$    & $\brkkkzlimit$     \\  \hline
Events & & \\
Data                  & \ndatc           & \ndatd             \\
Estimated background  & $\brkkkbkgd$    & $\brkkkzbkgd$           \\
\hline
Selection efficiency  & ($\kkkeffcc$)\%  & ($\kkkeffdd$)\%    \\
\hline
\end{tabular}
\end{center}
\end{table*}

The same criteria are used to select \taukkk and \taukkkz
decays except that the charged track is required to be a kaon.
The numbers of events are given in Table~\ref{table:kaon:results}
and found to be consistent with the estimated background prediction.
The background is almost entirely due to cross feed of \taupkk 
and \taupkkz decays and very little background from \qqbar events. 
The branching fractions are determined for each channel independently
and used to place upper limits on the branching fractions of
\begin{eqnarray*}
B(\taukkk) <  & \brkkklimit \\
B(\taukkkz) < & \brkkkzlimit 
\end{eqnarray*}
at the 90\% confidence level.

The \taukkzkz and \taukkpkm branching fractions are also 
predicted to be equal assuming isospin symmetry.
The \taukkpkm branching fraction is  $(2.1 \pm 0.8) \times 10^{-5}$
\cite {pdg} based on measurements from \babar\ \cite{babar:pkpkm} and
Belle \cite{belle:pkpkm}.
\babar\ finds that a \taukphi contribution can account for all of
the \taukkpkm decays.
This suggests that the \taukkk and consequently, the \taukkkz branching 
fractions should be small in the limit of isospin symmetry.

In summary, we have measured the branching fractions of the \taupkk and 
\taupkkz decays to be $\ba$ and $\bb$, respectively.
The \taupkk and \taupkkz decays can be modeled using \taukstarkz,
and \taukstarkzpz and \taukstarkzpim decays, respectively.
The \taupkk branching fraction is a significant improvement on the 
previous measurements and the \taupkkz  branching fraction is the first 
measurement.
In addition, we place the first upper limits on the branching fractions of 
$\brkkklimit$ and  $\brkkkzlimit$ on the \taukkk and \taukkkz decay modes
at the 90\% confidence level.

\begin{acknowledgments}
We are grateful for the 
extraordinary contributions of our \pep2\ colleagues in
achieving the excellent luminosity and machine conditions
that have made this work possible.
The success of this project also relies critically on the 
expertise and dedication of the computing organizations that 
support \babar.
The collaborating institutions wish to thank 
SLAC for its support and the kind hospitality extended to them. 
This work is supported by the
US Department of Energy
and National Science Foundation, the
Natural Sciences and Engineering Research Council (Canada),
the Commissariat \`a l'Energie Atomique and
Institut National de Physique Nucl\'eaire et de Physique des Particules
(France), the
Bundesministerium f\"ur Bildung und Forschung and
Deutsche Forschungsgemeinschaft
(Germany), the
Istituto Nazionale di Fisica Nucleare (Italy),
the Foundation for Fundamental Research on Matter (The Netherlands),
the Research Council of Norway, the
Ministry of Education and Science of the Russian Federation, 
Ministerio de Ciencia e Innovaci\'on (Spain), and the
Science and Technology Facilities Council (United Kingdom).
Individuals have received support from 
the Marie-Curie IEF program (European Union) and the A. P. Sloan Foundation (USA). 
\end{acknowledgments}


\end{document}